# A Relay-Chain–Powered Ciphertext-Policy Attribute-Based Encryption in Intelligent Transportation System

Aparna Singh[a], Geetanjali Rathee[a], Chaker Abdelaziz Kerrache[b] and Mohamed Chahine Ghanem[c,*]

[a]*Department of Computer Science and Engineering, Netaji Subhas University of Technology, New Delhi, India*
[b]*Laboratoire d'Informatique et de Mathématiques, Université Amar Telidji de Laghouat, Laghouat, Algeria*
[c]*Cybersecurity Institute, Department of Computer Science, University of Liverpool, Liverpool, UK*

## ABSTRACT

The very high growth of Intelligent Transportation Systems (ITS) has generated an urgent requirement for secure, effective, and context-aware data sharing mechanisms, especially over heterogeneous and geographically dispersed settings. This work suggests a new architecture that combines a relay chain–driven encryption system with a modified Ciphertext-Policy Attribute-Based Encryption (CP-ABE) scheme to tackle the double impediment of dynamic access and low-latency communication. The model proposes a context-aware smart contract on a worldwide relay chain that checks against data properties, including event type, time, and geographical region, to specify the suitable level of encryption policy. From such relay-directed judgment, On-Board Units (OBUs) encrypt data end-to-end by utilising CP-ABE and store ciphertext inside localised regional blockchains, preventing dependence on symmetric encryption or off-chain storage. High-sensitivity events are secured with firm, multi-attribute access rules, whereas common updates use light policies to help reduce processing burdens. The crypto system also adds traceability and low-latency revocation, with global enforcement managed through the relay chain. This distributed, scalable model provides a proper balance between responsiveness in real time and security and is extremely apt for next-gen vehicular networks that function across multi-jurisdictional domains.

## 1. Introduction

The Internet of Vehicles (IoVs) is a complete network system that combines numerous technologies, including electronic sensing, data transmission, and advanced information technology [1]. It allows for a wide range of communication and information sharing by fusing the automobile with the Internet. This improves road use efficiency, reduces traffic accidents, and improves traffic management in addition to allowing cars to autonomously gather, process, and share traffic information. IoVs can be used for a variety of purposes, including encouraging the creation of intelligent transportation networks and improving individual driving experiences [2]. The increasing number of smart cars is anticipated to generate and interchange massive amounts of data due to the quick development of vehicular applications and services, and the network traffic that has to be managed will be immense. However, using typical cloud-based storage and management directly will also present significant challenges for the IoV's high mobility, low latency, context complexity, and heterogeneity features. Strong interoperability and compatibility across IoV entities from various service providers are similarly challenging to guarantee. To accommodate future IoV development and effectively utilise the promise of Intelligent Transport System, ITS, the data interchange and storage infrastructure for IoV must be decentralised, distributed, interoperable, flexible, and scalable. Numerous studies [3], [4], and [5] have designed distributed vehicular network systems to address the aforementioned issues. For the computation results to be consistent, the distributed system requires that all nodes work together to communicate real-time data. Data security, data availability, and node privacy are some of the issues that arise during the data sharing process. First off, vehicles cannot completely trust data uploaded from roadside unit facilities to take part in the data-sharing program because of the potential for a single point of failure (SPOF). Second, malevolent nodes could alter data or spread false information to deceive other nodes and cause system errors. Thirdly, the creation of information-isolated islands might be the outcome of accomplishing complete data sharing between cars via point-to-point transmission. Lastly, the absence of an authorisation method raises the possibility of data availability issues and car privacy data leaks.

### 1.1. Motivation

Blockchain is a new distributed ledger system that relies on decentralisation, non-tampering, and data traceability to accomplish extraordinary development and breakthroughs in research and application. Numerous survey studies have already addressed the applications of blockchain in ITS as well as its integration into vehicle networks, both with and without cryptographic approaches [6], [7], [8] and [9]. However, most blockchain-based IoV applications use RSUs as communication media and consensus executors [10], [11], and [12]. Vehicles can only use RSUs to communicate or use them as the certificate authority for vehicular identity registration and revocation, which can fail to access services because of RSUs' SPOF and raise the threat of adversary corruption. The adversary can forge RSU identity to get vehicle information or take control of message transmission. In addition, as roadside infrastructure, RSUs cannot provide complete coverage in sparsely populated mountainous regions and other locations; thus, IoV applications that are entirely dependent on RSUs cannot perform data transmission in uncapped areas. The ABE concept was first proposed by Sahai et al. [13] [14]. From its introduction, it gained extensive interest and has become a very useful cryptographic building block with tremendous potential uses. Consequently, the ABE scheme's design has attracted interest from cryptographers across the globe. Researchers have made numerous efforts in making improvements in the trade-offs between security, efficiency, expressiveness, and scalability. These endeavours have resulted in a series of outstanding contributions [15], [16], [17], [18], [19] and [20] to the ABE field.

[*]Dr Mohamed Chahine Ghanem is the corresponding author.
✉ aparna.singh.phd21@nsut.ac.in (A. Singh);
geetanjali.rathi@nsut.ac.in (G. Rathee); ch.kerrache@lagh-univ.dz (C.A. Kerrache); mohamed.chahine.ghanem@liverpool.ac.uk (M.C. Ghanem)
ORCID(s):

### 1.2. Contribution

Drawing upon the lessons from similar existing work, we present a federated ITS data-sharing system that integrates a relay-chain with Ciphertext-Policy Attribute-Based Encryption (CP-ABE) scheme (RC-CP-ABE) in ITS. In our system, OBU encrypts its sensor's data directly with CP-ABE under policies made in real-time by a relay-chain smart contract. Low-sensitivity updates skip encryption to save latency, while critical events trigger strict, multi-attribute access controls. Local blockchains store the resulting ciphertexts and metadata, and the relay chain stores global attribute definitions, revocation lists, and cross-region indexing to provide efficient inter-jurisdictional access. Our adapted CP-ABE makes use of light operations and inherent traceability, weaving each user's encrypted identity into the secret key and supporting efficient revocation based on the q-SDH assumption. The following are the main contributions of this paper:

1. Relay-Guided Context-Aware Encryption Switching: We present a new smart-contract-based mechanism on a relay chain that dynamically controls CP-ABE policy strictness and whether to even encrypt data, depending on event sensitivity (e.g. high-security for accidents, low-latency raw broadcasts for mundane updates).

2. Consistent Ciphertext Structure with Dual-Mode Payload: Our case maintains a single $CT_f$ structure where OBUs insert either CP-ABE–encrypted or plain data under a light policy framework, easing storage, parsing, and decryption processes for all regional chains.

3. In-Built OBU-Only Encryption: We prove that low-capability OBUs are capable of performing case-light CP-ABE operations for compact ITS messages, supporting genuine end-to-end confidentiality without depending on RSU offload or hybrid AES solutions.

4. Federated Multi-Region Chain Architecture with In-Chain Integration: We introduce a two-layer blockchain system—regional chains for zone-specific storage and a relay chain for global policy coordination and cross-region indexing—directly connected (no gateways) with low-latency inter-chain communication and uniform revocation.

Furthermore, the remaining sections of this paper are structured as follows: Section 2 presents a brief review of related work focusing on IoV and ITS. In Section 3, the proposed framework is outlined in detail, including the system architecture and the background cryptographic base supporting the modified CP-ABE. Sections 4 and 5 examine the security aspects of the proposed approach and provide a comparative analysis with existing models in terms of computational cost and storage overhead, and latency in the Ethereum network.

## 2. Related Work

This section presents the number of schemes and approaches presented by several researchers and scientists. Table 1 presents the overall discussion of proposed approaches discussed by existing researchers.

CP-ABE is a potent instrument for fine-grained access control in data-sharing systems, particularly in security-critical fields like cyber-physical systems and the ITS. Its capacity to impose expressive access policies directly on the ciphertext enables data owners to maintain control over what data is accessible to whom, without depending on perpetual connectivity or a storage provider's trust. Routray et al. [21] introduced a lightweight, pairing-free CP-ABE scheme for cloud-assisted Cyber-Physical Systems (CPS) to overcome the latency and computational overhead concerns in conventional CP-ABE. The scheme utilises elliptic curve point multiplication and takes advantage of fog computing to shift the computationally expensive encryption task, thereby decreasing device-side load as well as enhancing performance. Working on policy hiding, Zhao et al. [22] suggested an entirely policy-concealed CP-ABE scheme for secure data sharing over CPS, based on an LSSS-based access structure. It avoids leakage of sensitive access policies by re-expressing them as blinded policy sets and hence maintains privacy for users. A similar scheme was introduced by Cui et al. [2], who resolved single-point failure in their work through dispersing trust across multiple attribute authorities that create intermediate keys. Extending the work on policy hiding, Zhao et al. [24] in their article introduced a publicly auditable data-sharing scheme for VANETs based on an expressive CP-ABE algorithm complemented with white-box traceability and partial policy hiding. To ensure prevention from key misuse, it utilises two cooperating authorities to produce user keys, which allow public identification of key leakers. Fog nodes were used to support computations with large loads, enabling the system to become practical for resource-limited vehicles.

Nevertheless, standard CP-ABE schemes usually use a single trusted authority to delegate user secret keys, which creates a central point of failure, a bottleneck on scalability, and susceptibility to authority compromise. To address this, researchers have introduced Multi-Authority CP-ABE (MA-CP-ABE) models, in which disjoint attribute sets are governed by multiple independent authorities. Datta et al. [25] presented a MA-CPABE scheme where any party may function as an authority without global coordination, except for the setup phase. It offers security against adaptive key queries as well as compromise of authorities over time. Later, Yao et al. [26] presented an improved version of it by integrating lattice-based cryptography, a global ID model, a two-stage sampling algorithm, and monotonous linear secret sharing schemes (M-LSSS). Contrary to regular ABE, it enables several authorities to issue keys separately, thus removing single-point failure and overload problems. Horng et al. [27] introduced another MA-CP-ABE scheme designed for vehicular ad hoc networks (VANETs) secure data sharing. It solves important issues such as user privacy, access control, and revocation, advancing the limitations of single-authority CP-ABE schemes. The method offloads burdensome encryption and decryption processes to cloud nodes to lighten the computation burden on resource-constrained in-vehicle units (OBUs). Attributes are decentralized and controlled by multiple service providers, supporting fine-grained access control and scalability. Meng et al. [28] proposed a dual hybrid ciphertext-policy attribute-based encryption (DH-CPABE) scheme for vehicular opportunistic computing (VOC) to support secure, fine-grained sharing of data without a trusted authority (TA).

**Table 1**
Comparative Table of Related Works and Research Gaps

| Author Name | Proposed Mechanism | Application | Survey | Performance Analysis | Limitation |
| --- | --- | --- | --- | --- | --- |
| Routray et al. [21] | lightweight, pairing-free CP-ABE scheme | CPS | × | latency and computational overhead | key management overhead |
| Zhao et al. [22] | LSSS-based access structure | CPS | × | sensitive access policies | storage overhead |
| Do et al. [23] | SHAP algorithm | Electric vehicle | × | Anomaly detection | Latency |
| Cui et al. [2] | dispersing trust across multiple attribute | Electric vehicle | × | work on policy hiding | communication overhead |
| Zhao et al. [24] | data-sharing scheme for VANETs | VANETs | × | large loads | Latency in automation driving |
| Datta et al. [25] | MA-CPABE scheme | ITS | ✓ | security against adaptive key queries | computation overhead |
| Yao et al. [26] | lattice-based cryptography, a global ID model | vehicle | ✓ | secret sharing schemes | delay in real-time communication |
| Horng et al. [27] | MA-CP-ABE scheme | VANET | × | single-authority CP-ABE | storage overhead |
| Meng et al. [28] | dual hybrid ciphertext-policy attribute-based encryption | ITS | × | reduced computations, storage | long delays |
| Chen et al. [29] | data security cross-chain framework | IoV | × | fair data verification | delay |
| Wang et al. [30] | PoS-PBFT blockchain-based data sharing | IoV | × | lightweight blockchain consensus | computation overhead |
| Gao et al. [31] | CP-ABE (OHP-CP-ABE) | VANETs | × | defending against risks | long delay |
| Jiang and Lv [32] | zero-knowledge proof (ZKP)-enabled distributed | CPS | × | safe symmetric key management | key management overhead |
| Singh et al. [33] | blockchain-based decentralised trust management | IoV | × | enhance transaction throughput | long delay |

Encouraging the shortcomings of traditional access control systems, blockchain has been combined with CP-ABE to deliver decentralised, tamper-evident data management in IoV and CPS systems. The fusion of smart contracts and CP-ABE allows smart enforcement of access policies, open auditing, and safe data exchange independent of a central authority. Such a combination solves the issues of trust, data origin, and revocation in decentralised vehicular scenarios. As interactions among data in IoV grow more complicated city by city and region by region, recent research has pushed such blockchain-based CP-ABE frameworks into cross-chain and multi-chain environments. Chen et al. [29] in their paper presented a data security cross-chain framework for urban IoV to support data confidentiality and system fragmentation. Hash time lock contracts are employed for inter-city data trading among same-brand and same-model vehicles, and trusted relay chains for intra-city trading among diverse models. The designed HR-CP-ABE algorithm extends CP-ABE by adding hierarchical decryption, revocation-supporting vehicle attributes, and time-controlled access to sensitive data. Wang et al. [30] in this paper, proposed a hybrid PoS-PBFT blockchain-based data sharing scheme IoV that enabled vehicles and RSUs to jointly participate in lightweight blockchain consensus, guaranteeing fair data verification and authorization even without RSUs. Gao et al. [31] introduced "TrustAccess", a blockchain-enabled access control system that maintains both policy and attribute privacy during safe data sharing based on an optimized hidden policy CP-ABE (OHP-CP-ABE). It avoids the use of intermediaries, thus decreasing trust establishment expenses and defending against risks such as single points of failure. The scheme provides policy privacy through concealed structures and accommodates a wide universe of attributes, while attribute privacy is maintained through the use of a multiplicative homomorphic ElGamal cryptosystem when generating keys. To solve security and scalability problems, Jiang and Lv [32] in this work, the authors puts forward a zero-knowledge proof (ZKP)-enabled distributed authentication protocol with Inter Planetary File System (IPFS). It employs the Schnorr protocol for vehicle privacy-preserving authentication and proxy re-encryption through RSUs for safe symmetric key management. The encrypted information is cached in IPFS servers, whereas Hyperledger Fabric keeps authentication logs and data indexes. Singh et al. [33] proposed a blockchain-based decentralised trust management system for the IoV to counteract security and privacy attacks from untrustworthy peers.

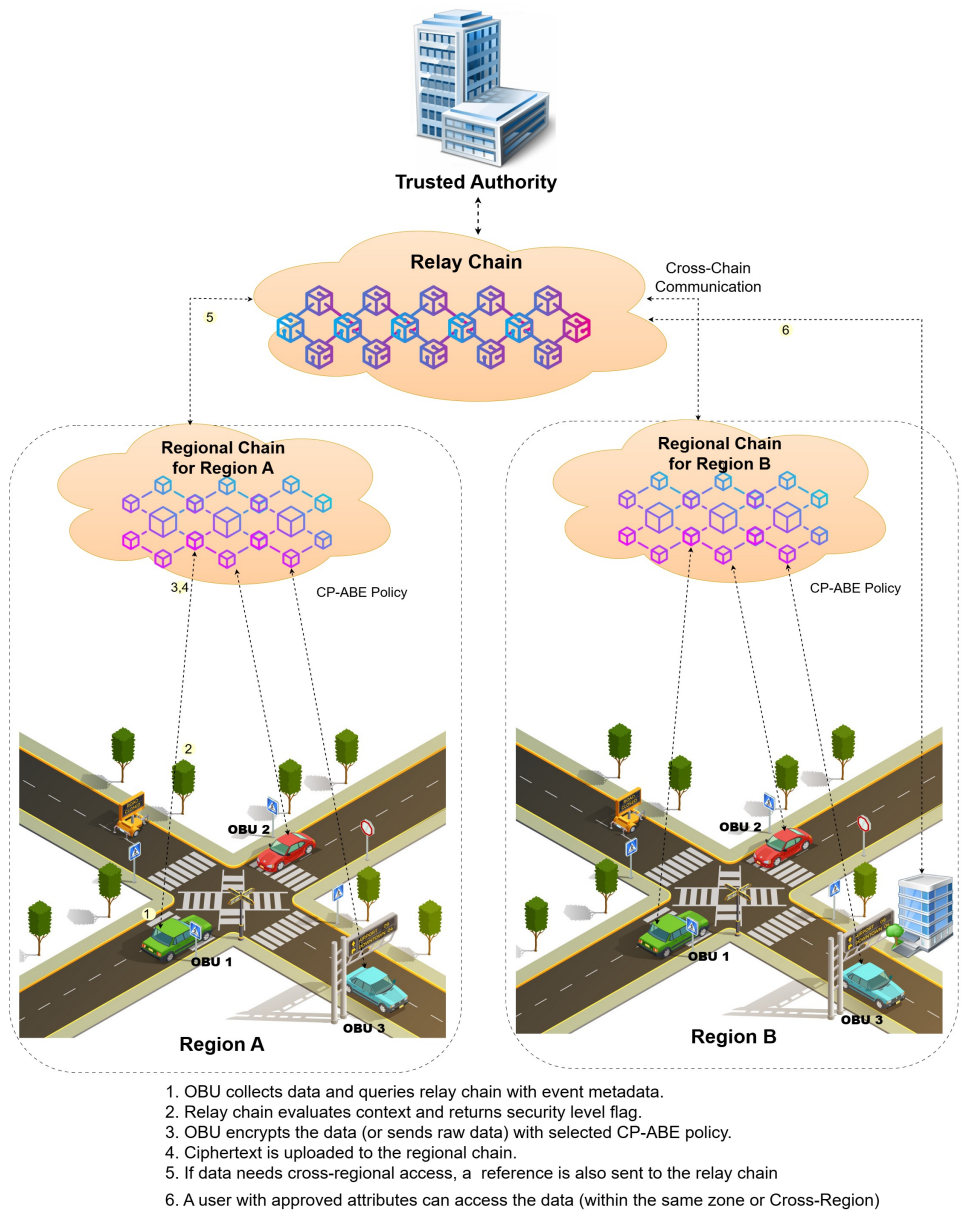

1. OBU collects data and queries relay chain with event metadata.
2. Relay chain evaluates context and returns security level flag.
3. OBU encrypts the data (or sends raw data) with selected CP-ABE policy.
4. Ciphertext is uploaded to the regional chain.
5. If data needs cross-regional access, a reference is also sent to the relay chain
6. A user with approved attributes can access the data (within the same zone or Cross-Region)

**Figure 1:** System Architecture

It employs smart contracts and adopts blockchain sharding to alleviate the main blockchain load and enhance transaction throughput [34–38].

## 3. Proposed System Architecture
### 3.1. System Architecture

The system model proposed utilises a relay chain-directed, attribute-based encryption structure optimised for secure and efficient data sharing in ITS. The model solves the issue of real-time data exchange and fine-grained access control across geographically distributed regions with lightweight operations compatible with resource-limited devices like On-Board Units (OBUs). The basic building blocks of the model, as illustrated in the Figure 1, are classified as follows:

***On-Board Unit (OBU)***: There is an OBU in each vehicle that senses and produces traffic-related information like accident warnings, vehicle speed, environmental conditions, etc. OBUS encrypt data in real-time with a customised variant of ciphertext-policy attribute-based encryption (CP-ABE) and sends encrypted data to the closest regional blockchain directly without a middleman or any external storage. OBUs also request the relay chain to decide on the encryption level depending on the context of the data.

***Roadside Units (RSUs)***: RSUs are enhancement nodes supporting OBUs with communication, routing, and query forwarding. Where OBUs are not resource efficient or are temporarily not connected, RSUs can act in place of OBUs to request the relay chain and encrypt, if authorised.

***Regional Blockchain***: Each geographical area (e.g., Region A, Region B) has its own unique blockchain ledger for keeping the encrypted traffic data produced by the OBUs in the area. These blockchains are tasked with ensuring persistence, local access control, and traceability of the data. The regional chains keep the encrypted data (in the form of CP-ABE ciphertext), its metadata (event type, location, timestamp), and the corresponding access policy.

***Relay Chain***: The relay chain serves as a worldwide controller linking all regional blockchains. It controls a registry of attribute definitions, access templates, user revocation lists, and context-aware encryption policies. It also contains a smart contract that, upon query by an OBU or RSU, provides a security flag stating whether the data must be encrypted with high-security or low-security policies, depending on contextual inputs like event type, region, or time.

*Trusted Authority (TA)*: A federated or centralised organisation that is involved in the CP-ABE scheme's initial configuration, i.e., creating the public parameters and master secret key. The TA also provides private attribute keys to compliant users like traffic police, municipal officials, and emergency rescue teams, depending on their roles and permissions.

*Authorised Users*: Organisations like traffic management centres, police officers, or city planners who are permitted to decrypt the encrypted data, provided their attribute keys meet the policy requirements embedded within the ciphertext.

The communication starts when an OBU retrieves traffic information. Before encryption, the OBU prompts the relay chain smart contract with metadata like event type, region, and timestamp. The smart contract executes this input against a globally defined context-sensitive policy template and responds with a boolean flag specifying the requisite security level. For instance, if the information relates to a night accident in a sensitive Region (e.g., Region A), the context is considered sensitive and the flag becomes true. Here, the OBU uses a high-security CP-ABE policy demanding rigorous access control (e.g., attributes Role = TrafficPolice AND Region = RegionA), along with encrypting the data before transmitting. On the other hand, if the information is a weather report or an ordinary traffic pattern update in a low-risk Region, the situation is non-sensitive and the flag is false. The OBU then uses a less restrictive policy like granting access with one attribute (e.g., Role = Analyst) and no added encryption of the data. This hybrid method promotes effective management of real-time traffic data while allowing strict access control for important events. To facilitate cross-regional access to data, the system separates locally important and globally important data. Encrypted ciphertext is, by default, kept in the regional blockchain where the OBU happens to be located. But if the relay chain finds that the data is of wider importance or that access can be needed by users in other regions, then the ciphertext—or a secure pointer to it—is also stored on the relay chain. This guarantees that permitted users beyond the originating region can find and retrieve the ciphertext to decrypt. The relay chain, therefore, serves not just as a policy coordinator but also as a federated index and access gateway for inter-region data sharing, maintaining both accessibility and access control integrity. After the policy decision, the information is encrypted through the enhanced CP-ABE scheme and kept in the regional blockchain with its metadata. If an authorised user wants to access the information, they inquire in the regional chain about the encrypted record. Only if their attribute keys match the policy embedded in the ciphertext can they decrypt it. Cross-regional access verification and revocation enforcement across all the Regions can also be served by the relay chain.

### 3.2. Assumptions Made

**Assumption 1: Decisional Bilinear Diffie-Hellman (DBDH):**

Let $G$ be a bilinear group of prime order $p$, with $g$ as its generator. Suppose that values $i, j, k \in \mathbb{Z}_p^*$ are randomly chosen. If an adversary is provided with the tuple $(g, g^i, g^j, g^k)$, it is computationally hard for the adversary $\mathcal{A}$ to distinguish whether a given element in the target group $G_T$ is of the form $e(g,g)^{ijk}$ or just a randomly chosen element from $G_T$. This assumption ensures the hardness of distinguishing meaningful bilinear map outputs from random values.

**Assumption 2: q-Strong Diffie-Hellman (q-SDH):**

Let $x \in \mathbb{Z}_p$ be selected at random, and let $g_1 \in G_1$, $g_2 \in G_2$. The q-SDH problem provides the tuple $(g_1, g_1^x, g_1^{x^2}, \ldots, g_1^{x^q}, g_2, g_2^x) \in G_1^{q+1} \times G_2^2$. The challenge is to compute a pair $(c, g_1^{1/(x+c)})$ for some $c \in \mathbb{Z}_p$. It is assumed to be infeasible for any adversary to generate such a pair efficiently, thereby ensuring the cryptographic strength of systems relying on this assumption.

### 3.3. Cryptographic Module

The cryptographic basis of the system proposed is constructed based on a customised Ciphertext-Policy Attribute-Based Encryption (CP-ABE) scheme tailored for lightweight and traceable vehicular network encryption. The scheme facilitates fine-grained access control by enabling the data owner (in this case, the OBU or RSU) to specify an access policy over attributes, whereby only users whose attribute keys meet the policy can decrypt the ciphertext. The CP-ABE scheme is built with four algorithms: Setup, KeyGen, Encrypt, and Decrypt. The Setup algorithm is executed by the trusted authority to obtain the public parameters (PP) and master secret key (MSK). The KeyGen algorithm inputs the MSK and the user's attribute set to generate an exclusive secret key. The Encrypt algorithm enables the data owner to set an access policy over attributes and encrypt the data in line with the policy. The Decrypt algorithm verifies that the attributes of the user meet the policy and, upon verification, retrieves the plaintext.

Here, encryption is directed by a smart contract that's relay-chain-enabled and dynamically chooses the policy strength. This process, known as "relay-guided encryption switching," decides whether high-security or low-security policies should be used during data encryption. High-security policies demand conjunctions of multiple attributes (e.g., Role = TrafficPolice AND Region = RegionA), whereas low-security policies could allow access with fewer requirements (e.g., Role = Analyst). This flexibility harmonises performance with confidentiality. In addition, the improved CP-ABE scheme incorporates identity traceability by associating encrypted data with traceable parameters (e.g., pseudonym or user ID) to allow each decryption to be traced back to an authenticated entity. The scheme also accommodates effective revocation through the storage of revoked attribute IDs in the relay chain smart contract for timely policy enforcement throughout different regions. Overall, the cryptographic module increases trust, security, and scalability in intelligent transport environments by integrating decentralized storage with dynamic access control.

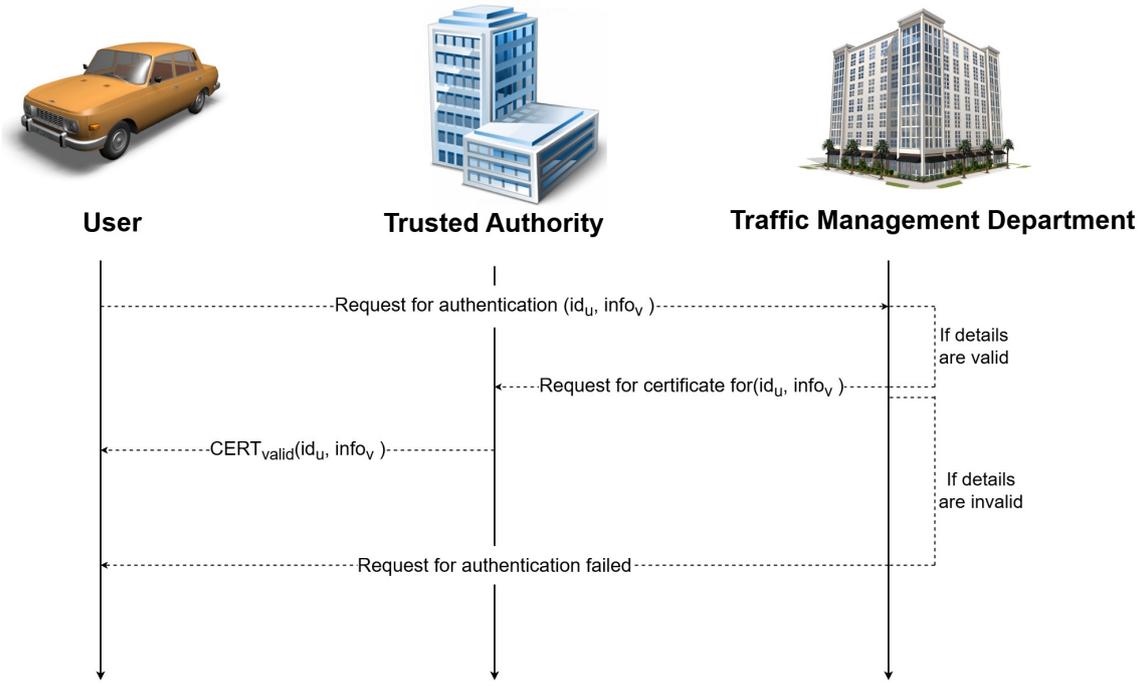

**Figure 2:** Process flow for user registration and authentication

*Global_Setup Phase*: The TA takes $(1^\lambda)$ as the input and initialises the *Global_Setup* algorithm to produce $PP_{TA}$ and $MSK_{TA}$ as output. Algorithm 1 elaborates on the generation of $PP_{TA}$ and $MSK_{TA}$.

---

**Algorithm 1:** Global Setup Algorithm

**Data:** $1^\lambda$
**Result:** PP and MSK
1. Execute $G(1^\lambda) to \to (G, G_T, p, e)$ ;
2. Select a generator $g \in G$ and random numbers $a, \varphi \in Z_p^*$;
3. Randomly select $\kappa_1 \in \mathcal{K}$, where $\mathcal{K}$ is available key space ;
4. Calculate $f = g^\varphi, X = e(g, f)^a, X_0 = e(g, f), h = g^\lambda$;
5. Choose two hash functions: $H\{0, 1\}^* \to Z_p^*$ and $H_1\{0, 1\}^* \to \mathcal{K}$ ;
6. Calculate $MSK_{TA} = (a, \varphi, \lambda, \kappa_1)$;
7. Store PP in $Relay\_Chain$;
8. Return $PP_{TA} = (g, h, f, X, X_0)$;

---

*Registration Phase*: Figure 2 represents this vital phase, needed for the authentication of the identity of the users involved in the communication, within this framework. Before the KeyGen process, each user has to apply for a trusted certificate to authenticate the identity of the user. The user will have to submit their government-issued id and the vehicle information $(id_u, info_v)$ to a traffic management department, which will verify the information given by the user. Once verified, the department will request the TA to issue a certificate $CERT_{valid}(id_u, info_v)$ for the verified user. With this valid certificate, the user can connect to the system and transmit information and also receive any information within the framework.

*Key Generation*: For each verified user, TA initiates the KeyGen algorithm by accepting each user's $CERT_{valid}$, which contains $id_u$ and their attribute set $A_s$. Each entry of $A_s$ is denoted by $\{\varkappa_1, \varkappa_2, ....\varkappa_k\} \subseteq Z_p^*$. Randomly select $\iota, \mu, \vartheta, v', v'', j', j'', j''', j'''' \in Z_p^*$. The TA computes, $\Upsilon = Sym\_Enc_{\kappa_1}(id_u)$. The private and public keys are calculated as:

$K_1 = g^{\frac{a-\alpha \iota}{\lambda+\Upsilon}}$, $K_2 = \Upsilon$, $K_{3,i} = g^{\alpha \iota (v_i + \varphi)^{-1}}$, $K_4 = \vartheta$,
$\Omega_1 = (K_1^\vartheta)^{v'}$, $\Omega_2 = X_0^{j'''}$, $\Omega_{3,i} = (K_{3,i}^\vartheta)^{j''}$, where $i \in |k|$

$$\Omega_4 = g^{v'}, \Omega_5 = f^{v''}$$

$SK_{id_u, A_s} = (K_1, K_2, K_{\{3,i\}}, K_4, v', v'', j', j'', j''', j'''')$

$PK_{id_u, A_s} = (\Omega_1, \Omega_2, \{\Omega_{3,i}\}, \Omega_4, \Omega_5)$

*Encryption*: The encryption is initiated when an OBU gathers contextual metadata like the nature of the event (e.g., accident, weather report), geographical area, and timestamp. The metadata is sent as a query to the relay chain's smart contract, which compares it against global policies. Depending on the result, the smart contract responds with a boolean flag that determines the encryption level. If the flag is true, the OBU understands that as a high-sensitivity context (for example, an accident has happened in a critical area at night) and chooses a high-security encryption policy with tighter access control requirements, commonly using several attributes like role and region. Conversely, if the flag is false, meaning a non-sensitive situation like a weather report or normal traffic status in a low-risk location, the OBU chooses a relaxed security policy. This policy may only need one attribute, like a general analyst role, to meet the access structure. After the policy has been established, the OBU goes ahead to encrypt using the modified CP-ABE scheme. A random $\beta \in G_T^*$ is selected, which is used to compute $key = H_1(\beta)$. Subsequent steps are elaborated below in the Algorithm 2.

---

**Algorithm 2:** Context-Aware Encryption Algorithm

**Data:** Data $Data$, access policy $(M, \rho)$, keyword $K_{word}$, metadata (event, region, time)
**Result:** Ciphertext $CT_f$ to be uploaded to Regional Chain

1. Query relay chain smart contract with metadata $(event, region, time)$;
2. Receive security flag from relay chain;
3. **if** *flag == True (High Security)* **then**
   4. Set policy $P \leftarrow$ strict (e.g., Role = TrafficPolice $\wedge$ Region = RegionA);
   5. Encrypt $Data$ using CP-ABE: $Enc_{data} = Sym\_Enc_{key}(data)$;
**end**
6. **else**
   7. Set policy $P \leftarrow$ minimal (e.g., Role = Analyst);
   8. Set $Enc_{data} \leftarrow Data$ ;
**end**
9. Select $\nu \in \mathbb{Z}_p^*$ and a vector $\upsilon = (\nu, \upsilon_2, ..., \upsilon_n)^T \in \mathbb{Z}_p^n$;
10. Compute $\nu_i = M_i \cdot \upsilon$ for each row $M_i$ in $M$;
13. Compute $CT_0 = \beta \cdot I^\nu$;
14. Compute $CT_1 = g^\nu$ and $CT_3 = h^\nu$;
15. Compute $CT_{3,i} = \frac{\rho(i)\nu_i}{\nu' \cdot H(K_{word})}$;
16. Compute $CT'_{3,i} = \frac{\nu_i}{\nu'' \cdot H(K_{word})}$;
17. Compute $CT_4 = X^{H(K_{word_0})} \cdot X^{\nu/H(K_{word})}$;
18. Construct $CT_f = (CT_0, CT_1, CT_2, CT_{3,i}, CT'3, i, CT_4, Encdata)$;
19. Upload $(M, \rho)$ and $CT_f$ to the appropriate regional blockchain;

---

***Trapdoor Generation***: The OBU of the user generates the keyword trapdoor $TKey_{word}$ using this process, by taking $SK_{id_u}$, $A_s$ and $Key_{word}$ as input. Initially, the user randomly selects $J, J_0 \in Z_p^*$ and calculates $TK_0 = J.(J')^{-1}$ and $TK_1 = \frac{J_0}{[J'H(Key_{word})]}$ and $TK_2 = K_2$. It further calculates $TK_3 = \upsilon_0 \cdot (J'')^{-1}$, $TK'_3 = JH(Key_{word}) \cdot (J'')^{-1}$ and $TK_4 = J_0 K_4$. Lastly, $TK_5 = J_0 K_4 \cdot H(Key_{word}) \cdot (J''')^{-1}$. All the computed components are added together to form the final keyword trapdor

$$TKey_{word} = (TK_0, TK_1, TK_2, TK'_3, TK_4, TK_5).$$

***Retrieval Phase***: Upon receiving the request from the user, the smart contract in the $Regional_{Chain}$ looks for the encrypted data within the blockchain network. The smart contract makes use of the Search Algorithm 3 described below to look for the data matching the keyword shared by the user, $TK'_{word}$. The algorithm ensures that the access policy, $(M, \rho)$ stored with the CT matches with the attribute set, $A_s$ of the user requesting the data. If the attribute set fails to align with the recorded access policy, the algorithm returns $\perp$. In the following algorithm, $P \subseteq [l]$ and is represented as a set $\rho(i)$ which belongs to $A_s$. j represents a group of constants where $j_i \in Z_{p_{i \in P}}$.

---

**Algorithm 3:** Search Algorithm

**Data:** $TK'_{word}$
**Result:** $\rightarrow CT / \perp$
1. Calculate

$$\Lambda = e[\Omega_4, \prod_{i \in X}(\Omega_{3,i})^{K_{3,i} \cdot j_i}] \cdot e[\Omega_5, \prod_{i \in X}(\Omega_{3,i})^{K'_{3,i} \cdot j_i}]$$

2. Calculate $\Gamma = e(\Omega_1, CT_1^{TK_2} CT_2)$;
3. Calculate $\Gamma_1 = \Gamma^{TK_1}$;
4. Calculate $\Lambda_1 = \Lambda^{TK_3}$;
5. **if** $\Omega_2^{TK_5} \cdot (\frac{\Gamma_1}{\Lambda_2}) = CT_4^{TK_4}$ **then**
   6
**else**
**end**
**else**
   7. return $CT_0, \Gamma^{TK_0}, \Lambda^{T'_K}, Enc_{data}$ to the user's OBU.
**end**
7. return $\perp$

---

***Decryption phase***: To decrypt the $Enc_{data}$ the OBU computes:

$$\beta = \frac{CT_0}{\Gamma^{TK_0}/\Lambda^{T'_K}{}^{1/uK_4}}$$

With access to the $\beta$, the OBU can compute $key = H_1(\beta)$ and subsequently calculate $Sym\_Dec(Enc_{data})$.

***Track phase***: This algorithm enables tracing of users who have accessed certain data in the past. The decryption key of every user has an included parameter $\Upsilon$, which incorporates the identity of the user $id_u$, making it easy to trace. Additionally, the identity is redundantly included in the $K_1$ parameter, which is obtained from $a$ and $\lambda$—two components of the MSK known only to the TA. This double embedding guarantees that users cannot hide their identities. This traceability function facilitates stronger user accountability and promotes transparency within the data-sharing process.

***Attribute updation***: Attribute updates necessitate the regeneration of both secret and public key pairs. This is due to the fact that key components like $K_3$ and $\Omega_{3,i}$ are obtained directly from the user's attribute set $A_s$. Whenever a user changes their attributes—either gaining or losing rights—the Trusted Authority (TA) needs to issue a new key pair based on the new set. For traceability reasons, the encrypted identity $\Upsilon$ is still included in the regenerated key. The relay chain also keeps a revocation list to avoid misusing outdated keys, allowing only correct and updated attributes to meet encryption policies. Figure 3 reflects the interaction between the core functions in the model described above.

## 4. Security Analysis

This section demonstrates both the security and traceability properties of the proposed scheme.

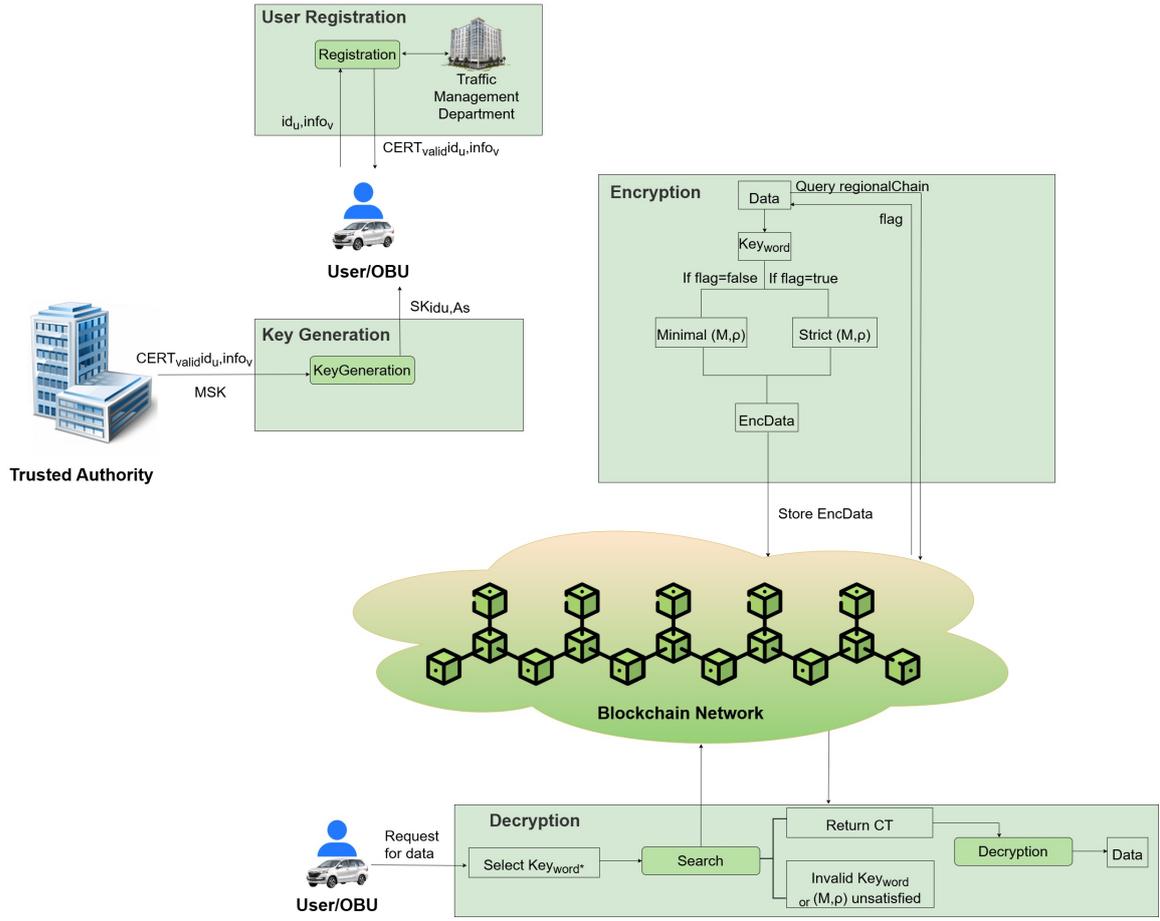

**Figure 3:** Detailed interaction between the core components of RC-CP-ABE

***Theorem 1***: Assuming the Decisional Bilinear Diffie-Hellman (DBDH) problem is hard, the proposed scheme achieves IND-CPA security.

***Proof***: To establish the security of our scheme, we define an adversary, **Adv**, and a challenger, **Ch**, who interact as follows. The adversary's objective is to compromise the system within polynomial time, while the challenger leverages the interaction to attempt to solve the DBDH problem. The security game proceeds as follows:

1. Initialization: The challenger **Ch** receives a tuple $(g, g^i, g^j, g^k)$ and either $e(g,g)^{ijk} \in G_T$ or a random element from $G_T$. Using these, **Ch** sets up the system's public parameters.

2. Query Phase: **Ch** answers all of $\mathcal{A}$'s queries regarding private key and trapdoor generation and returns the responses accordingly.

3. Challenge Phase: The adversary submits an access structure $(M, \rho)$, two equal-length messages $(m_0, m_1)$, and a pair of keywords $(KW_0, KW_1)$ to **Ch**. The challenger randomly selects $r \in 0, 1$, encrypts $m_r$ and $KW_r$, and sends the ciphertext back to $\mathcal{A}$.

4. Second Query Phase: **Adv** continues to make queries to **Ch** as before.

5. Guess Phase: **Adv** outputs a guess $r' \in 0, 1$. If $r' = r$, **Ch** outputs 'true'; otherwise, 'false'. If 'true',

the challenger can create a valid ciphertext with an advantage of

$$\frac{1}{2} + \epsilon,$$

where $\epsilon$ is the adversary's probability of successfully solving the DBDH problem. If 'false', the adversary fails.

According to Section 3.2, under Assumption 1, the scheme remains secure and infeasible to break within polynomial time.

***Theorem 2***: Provided the q-Strong Diffie-Hellman (q-SDH) assumption holds and $t' \geq t + t_e[O(|S|)\text{count}sk]$ (where count$sk$ is the number of secret keys, $t_e$ is the time for exponentiation, and $|S|$ is the number of attributes in set $S$), the scheme guarantees traceability.

***Proof***: If an adversary **Adv** can compromise the traceability of the scheme, then the challenger **Ch** can leverage **Adv** to solve the q-SDH problem. The traceability game is structured as follows:

1. Initialization: **Ch** is given $(g, g^i, g^j, g^k)$ and either $e(g,g)^{ijk} \in G_T$ or a random element from $G_T$. Using these, **Ch** constructs the system's public parameters.

2. Query Phase: The challenger responds to all secret key queries from **Adv**.

3. Challenge Phase: The adversary submits a modified secret key $SK^*_{id_u, A_S}$ to **Ch**. If $SK^*_{id_u, A_S}$ passes verification Algorithm 4, then **Adv** has succeeded in breaking traceability, allowing **Ch** to use $SK^*_{id_u, A_S}$ to address the q-SDH problem.

Since the q-SDH problem is assumed to be hard in polynomial time, **Adv** cannot feasibly produce a valid $SK_{id_u, A_s}*$ that would enable solving the q-SDH problem with non-negligible probability.

---

**Algorithm 4:** Key Verification Algorithm

**Data:**
**Result:** 0/1
1. The TA checks whether the

$$SK_{id_u}, A_s *$$

provided by **Adv** conforms to the structure:

$$SK_{id_u, A_s} *= (K_1, K_2, K_{\{3,i\}}, K_4, v', v'', j', j'', j''', j''')$$

where $K_1, K_{3,i} \in G$ and $K_2, K_4, j', j'', j''' \in Z_p^*$.
2. The TA then evaluates the validity of the following expression:

$$e(K_1, H \cdot g^{K_2})^{\mathcal{K}} \cdot e(f^\alpha, \prod_{i \in [\mathcal{K}]} (K_{3,i})) = X^{\mathcal{K}}$$

3. If the equation is satisfied, the algorithm outputs 1. Otherwise, it returns 0.

---

## 5. Performance Analysis

The execution and result calculation method made an exhaustive performance assessment of Ethereum-based smart contracts from a simulated testbed, using Replit []. Two primary Solidity contracts- *RelayChain* and *RegionalChain*, were deployed through Web3.js, set with realistic parameters like a 2-second block time and 20 Gwei gas price (standard). These were evaluated under three sample functions: evaluatePolicy() (view function), uploadCiphertext() (state-altering data uploading function), and revokeAttribute() (access revocation function). Their performance was tested using concurrent user testing, scaling from 5 to 20 users in steps, with ten iterations under each configuration to guarantee statistical consistency. Latency was measured with accurate JavaScript timing mechanisms, whereas gas consumption and transaction fees were calculated from receipts and estimates. The latency graphs in Figure 4 show the variation in EvaluatePolicy(), UploadCiphertext(), and RevokeAttribute() latencies as the number of users is increased. As illustrated, all three latencies increase smoothly as more users use the system concurrently. The evaluation latency increases moderately, showing that the attribute checking and policy matching procedures do not become too heavy as the number of users is increased. By contrast, upload latency rises more sharply, as would be expected with the increased computational and transactional overhead of adding new encrypted data or policies to the blockchain. Revocation latency also increases continuously, mirroring the increased

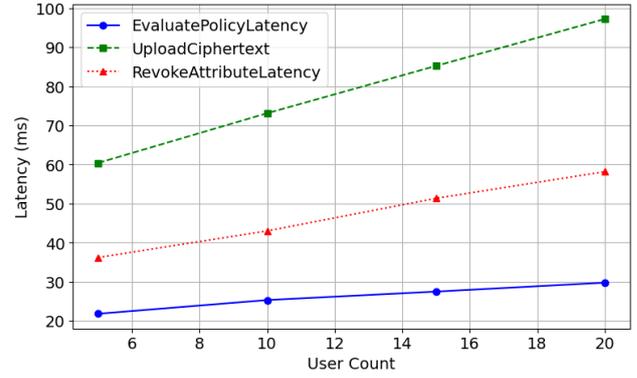

**Figure 4:** Latency comparison for three major functions

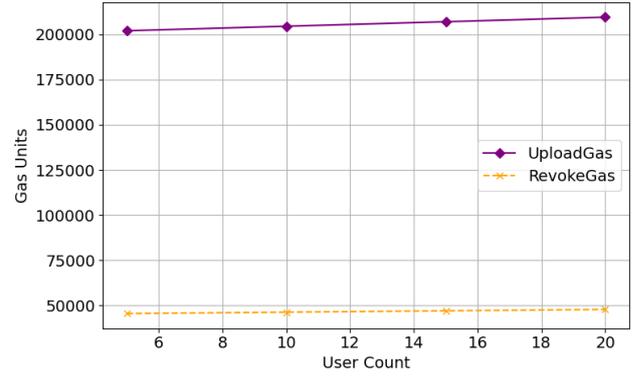

**Figure 5:** Gas utilisation on Ethereum for uploadCiphertext() and revokeAttribute() functions

effort to update access permissions and to disseminate revocation events.

### 5.1. Performance testing on Ethereum

The graphs in Figure 5 demonstrate the computational expense, expressed in Ethereum gas units, for the upload and revocation functions. Both measurements rise linearly with users, exhibiting stable and controllable scalability. UploadGas() is always higher, since uploading requires bigger state changes or more advanced smart contract logic. RevokeGas(), though lower, exhibits a similar pattern because contract updates need to be responsive to access control changes. We also calculated an efficiency ratio (millisecond latency per 1000 gas units) to assess how well every smart contract function leverages compute resources. The smaller the ratio, the more efficient the gas is. The deployment of both the smart contracts shows that although uploadCiphertext() uses more overall gas, it is more efficient at a ratio of 0.30-0.47 under user loads, as opposed to revokeAttribute(), 0.80-1.22. This implies revokeUser, while its total gas usage is lower, is less computationally efficient and would most gain from optimisation. The efficiency ratio also indicates that both functions become less gas efficient with higher user load, with efficiency worsening by about 55% when scaling from 5 to 20 concurrent users.

### 5.2. Performance of the modified RC-CP-ABE

This section evaluates the storage requirements of our model in comparison to existing schemes, focusing on four primary parameters: public parameters, user secret keys, and ciphertext size. We also analyse the computational overhead. Table 3 presents these storage metrics in bytes. For notation, let $|\mathcal{G}|$, $\mathcal{G}_T$, and $\mathcal{Z}_p$ denote the byte sizes of elements in groups $\mathcal{G}$, $\mathcal{G}_T$, and $\mathcal{Z}_p$, respectively; $T_1$, $T_2$, and $T_p$ represent the times for exponentiation in $G$, exponentiation in $G_T$, and bilinear pairing operations; and $\mathcal{A}_s$, $\mathcal{U}$, and $\mathcal{L}$ denote, respectively, the number of attributes in a user's set, the size of the universal attribute set, and the number of rows in the access-policy matrix $M$.

**Table 2**
Comparison based on Storage cost

| Ref | SK | PP | CT |
|---|---|---|---|
| Li et al. [39] | $(3 + |\mathcal{A}_S|)|\mathcal{G}| + 2|\mathcal{Z}_p|$ | $4|\mathcal{G}| + |\mathcal{G}_T| + 2|\mathcal{Z}_p|$ | $(2|\mathcal{L}| + 3)|\mathcal{G}| + |\mathcal{G}_T|$ |
| Zeng et al. [17] | $(\mathcal{A} + 2)2|\mathcal{G}|$ | $4|\mathcal{G}| + |\mathcal{G}_T|$ | $(3L + 2)|\mathcal{G}| + |\mathcal{G}_T|$ |
| Zhao et al. [24] | $(3|\mathcal{A}_s|)|\mathcal{G}| + 2|\mathcal{Z}_p|$ | $(3 + |\mathcal{U}|)|\mathcal{G}| + |\mathcal{G}_T| + 2|\mathcal{Z}_p|$ | $(2\mathcal{L} + 2)|\mathcal{G}| + |\mathcal{G}_T|$ |
| Proposed model | $(\mathcal{A}_s + 1)|\mathcal{G}| + 5|\mathcal{Z}_p|$ | $4|\mathcal{G}| + 2|\mathcal{G}_T|$ | $4|\mathcal{G}| + 2|\mathcal{G}_T| + 2L|\mathcal{Z}_p|$ |

**Table 3**
Comparison based on Computational Cost

| Ref | Key Generation Cost | Encryption Cost | Decryption Cost |
|---|---|---|---|
| Zeng et al.[17] | $(5\mathcal{L} + 2)T + 2T_T$ | $2T_P + 2\mathcal{A} + 1$ | $2T_P + (2\mathcal{A} + 1)T$ |
| Zhao et al. [24] | $(\mathcal{A}_s + 2)T_p + 2T + \mathcal{A}_s T_T$ | $2\mathcal{L}T$ | $T_P + 2T$ |
| Proposed model | $(\mathcal{A}_s + 1)T$ | $4T + 3T_T$ | $T$ |

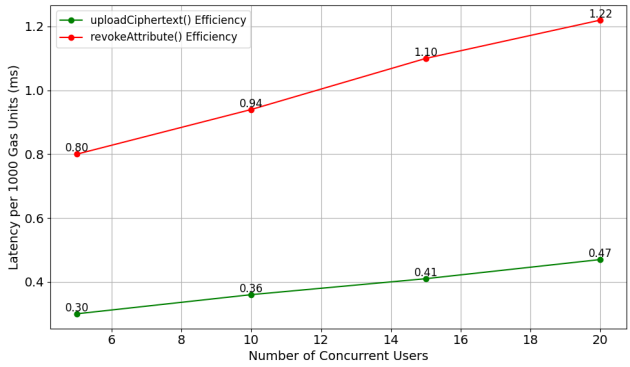

**Figure 6:** Comparison of uploadCiphertext() and revokeAttribute() efficiency in terms of latency per 1000 gas unitss

We used the Python-based Charm framework and Pairing-Based Cryptography, PBC Library-0.5.14 version, to develop the suggested strategy for bilinear cryptography. The experiment was conducted using a laptop with an Intel Core i5 processor running at 2.6 GHz and a Windows subsystem based on Ubuntu 22.04.00. For the implementation, an elliptic curve known as "SS512" that supports Type A pairing and has the statement $Equation: y^2 = x^3 + x$ over the finite field $F_p$ was selected. Due to its symmetric nature, any element in group $\mathcal{Z}_p$ takes up 160 bits, while elements in group $\mathcal{G}$ take up 1024 bits. The same is true for elements in groups $\mathcal{G}_T$. The computation time used by some of the similar techniques and ours is further detailed in Table 3. The computing time for exponentiation calculation in group G, $G_T$, and bilinear pairing is represented by the notations $T$, $T_T$, and $T_P$. According to the parameters listed above, $T = 8.005ms$, $T_T = 1.91ms$, and $T_P = 16.01ms$ for a processor with 8 GB of RAM.

The graphs in Figure 7 illustrate the comparison of the key generation, encryption, and decryption costs of three schemes, Zeng et al. [17], and Zhao et al. [24] and the proposed scheme, with the size of $A_s$ varying.

***Key Generation Cost:*** The Figure 8 (a) illustrates that the suggested model always needs fewer times to generate keys than the other two schemes. The reason is primarily that the proposed method scales linearly with the size of the attribute set, but with a lower constant factor. In contrast, Zeng et al.'s [17] scheme incurs more computationally expensive operations, incurring higher time expenses, particularly as the size of the attribute set increases.

***Encryption Cost***: The encryption time for the model proposed in the paper is almost independent of the size of the attribute set $A_s$, which indicates its light-weight nature. For Zeng et al. [17], encryption time is increasing with the size of $A_S$ because of more costly exponentiation and pairing operations. Zhao et al.'s [24] scheme has a linear increase, but still takes typically more time than the proposed method.

***Decryption Cost***: Decryption cost for the presented model is negligible and independent of the size of the attribute set, emphasising its suitability for resource-scarce systems such as IoT devices. Zeng et al.'s [17] decryption cost is linearly increasing with the size of the attribute set, whereas Zhao et al.'s [24] scheme is about constant but higher than the proposed model. Overall, these timing comparisons demonstrate that the proposed scheme is more computationally efficient in all major cryptographic operations, making it more feasible for real-time or large-scale applications.

The second group of graphs in the Figure 8 presents the storage cost—in terms of bytes—of three primary elements: the Secret Key ($SK_{id_u}, A_s$), Public Parameters (PP), and Cipher Text ($CT_f$) for four schemes: Li et al. [39], Zeng et al. [17], Zhao et al. [40], and the Proposed Model.

***Secret Key Storage***: The proposed model has the least storage overhead for secret keys at all attribute set sizes since its key size increases linearly with the number of attributes, but employs fewer elements. Zhao et al. [24] and Li et al. [39] take much more storage space because they have extra parameters and larger element sizes. Zeng et al.'s [17] key size also increases more rapidly and is indicative of a more intricate key structure.

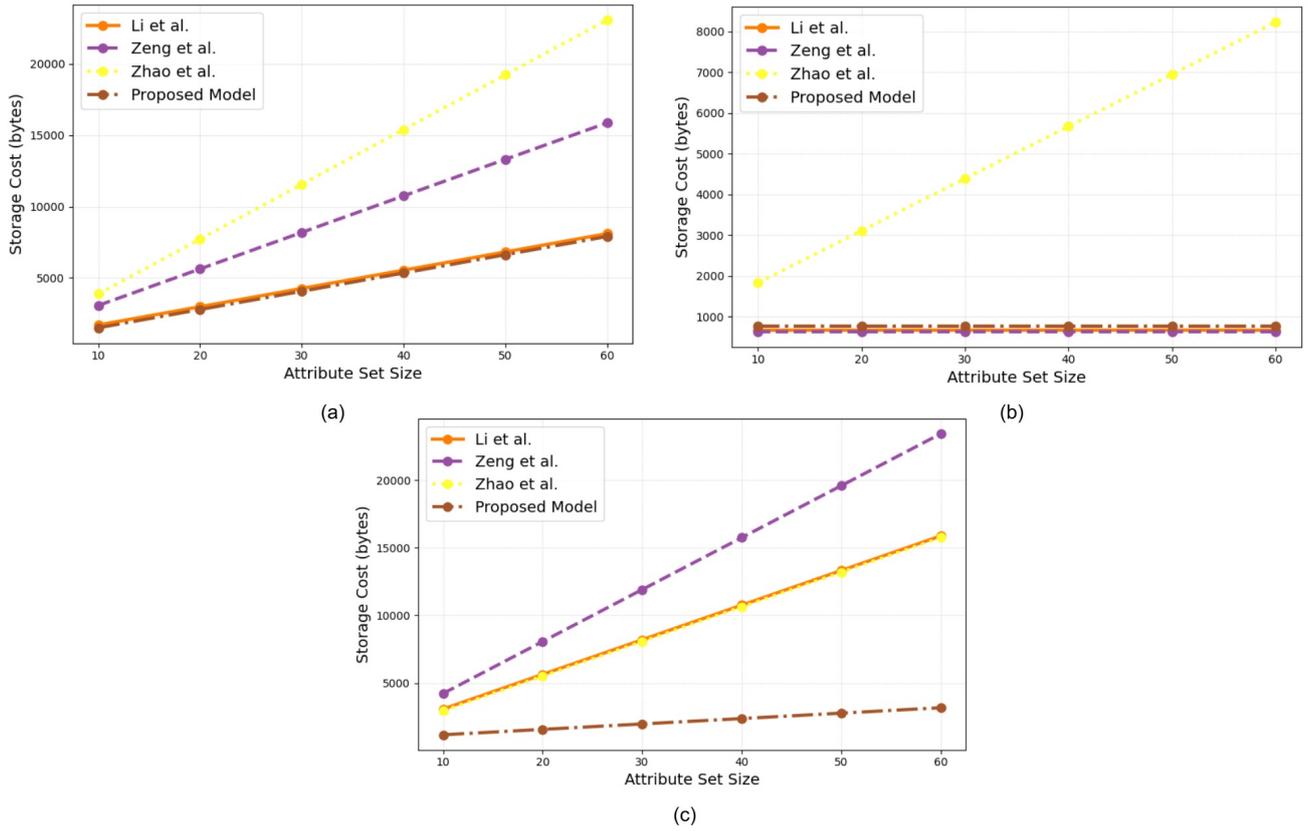

Figure 7: Computational cost comparison based on (a) Key generation (b) Encryption (c) Decryption

*Public Parameter Storage*: Public parameter size is relatively stable for the model proposed and for Zeng et al.[17], since their parameters are less dependent on attribute set size. Li et al. [39] and Zhao et al. [24] have increased and relatively growing storage costs owing to their parameter construction involving more elements connected with the attribute universe.

*Ciphertext Storage*: Ciphertext length goes up with attribute set size in all schemes. The suggested model, however, is able to keep the increase low by optimising the structure of the ciphertext, finding an equilibrium between pairing groups and attribute-dependent components. Zhao et al.'s [24] and Zeng et al.'s [17] ciphertext length increases at a higher rate owing to an increased number of attribute-dependent components. Li et al.'s [39] ciphertext length also goes up significantly with attributes.

The cost of storage comparison evidently emphasises the space efficiency of the given model, which is essential for settings such as IoT, where bandwidth and memory are scarce.

## 6. Conclusion

In this paper, we introduced a novel decentralised data sharing system suitable for ITS by combining an enhanced CP-ABE scheme with local blockchain and relay chains. By removing the centralised proxy server and utilising smart contracts, we improved the trust, transparency, and auditability of the system. The use of a decentralised ledger enabled us to provide efficient and secure off-chain storage of encrypted vehicular data with high immutability. The integrated function of a Trusted Authority simplified the initialisation and key distribution functions without overburdening the trust model. Our system further supports regional scalability via the insertion of relay chains as an option, providing equal security policies throughout distributed regions. Extensive simulation results illustrate that the presented model is superior compared to other schemes in computational efficiency and communication overhead. The computational cost of retrieving the secret key is significantly reduced in the proposed method. Similarly, no matter how big the attribute set is, the proposed scheme's encryption time remains constant at 37.75 ms and offers a notable 83% reduction in decryption time compared to other methods. In addition, smart contract usage enabled automatic revocation and attribute updates without third-party servers, reducing latency and improving responsiveness. Such results establish that the solution proposed not only satisfies the rigorous security and privacy requirements of ITS environments but also has concrete scalability and efficiency appropriate for real-world deployment. In the future, this model can be expanded further to include dynamic and real-time updates of various policy types where access structures change depending on context information like traffic flow, location, or environmental data, to enhance its usability in sophisticated urban IoV settings.

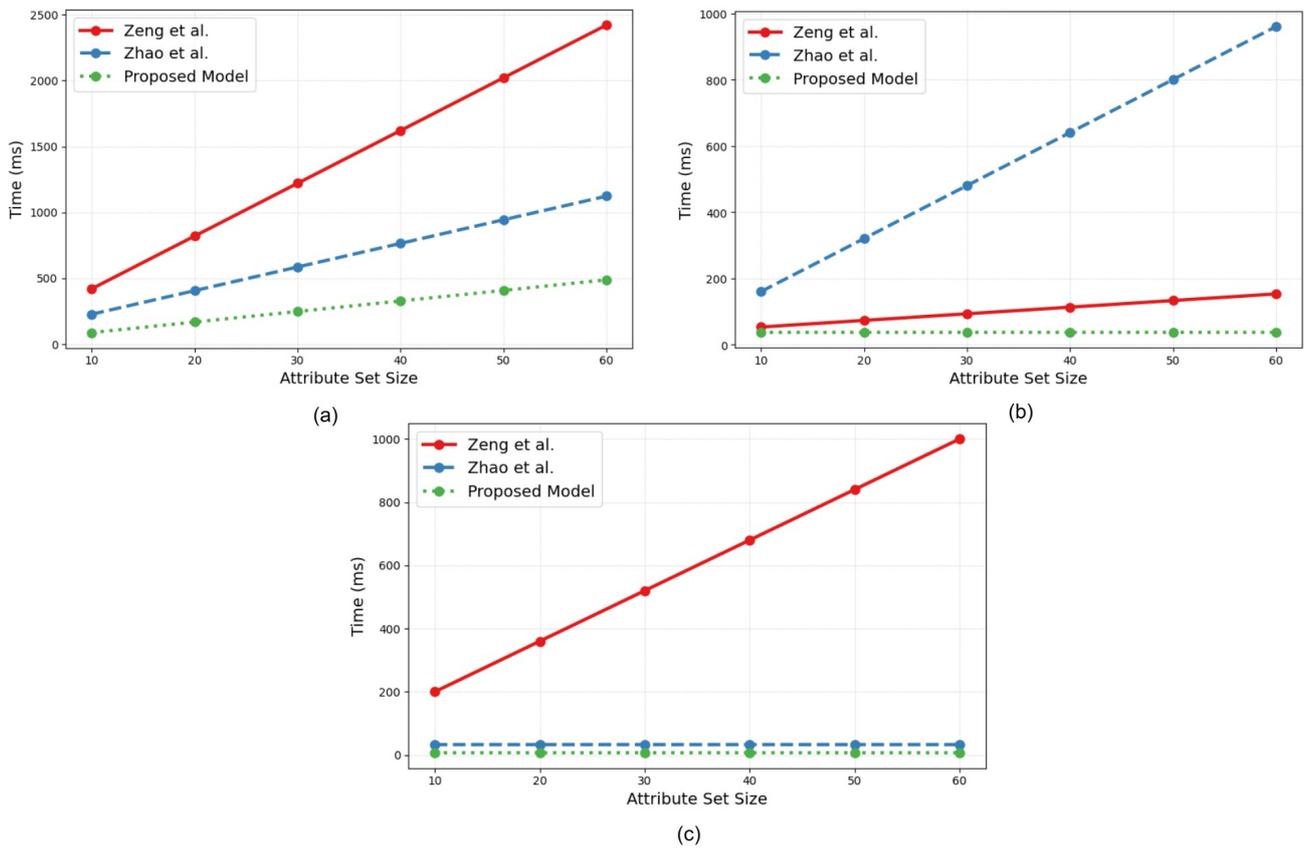

**Figure 8:** Storage cost comparison based on (a) Secret Key generation (b) Public Parameter generation, (c) Ciphertext generation

## Footnotes
### Acknowledgments
This work has been funded by R&D project SERB-SURE **SUR/2022/001051** in Netaji Subhas University of Technology. The Gold Open Access APC is supported by the University of Liverpool.

### Declaration of Competing Interest
No competing financial interests or personal affiliations are associated with the authors that could have impacted the content of this paper.

### Data Availability
The data supporting the findings of this study are available upon request from the corresponding author.

### Ethical Approval
This research was deemed not to require ethical approval.

## CRediT authorship contribution statement
**Aparna Singh:** Conceptualization; Methodology; Data Curation; Resources; Investigation; Visualization; Writing original draft; Writing, review, and editing. **Geetanjali Rathee:** Conceptualization; Methodology; Data Curation; Resources; Investigation; Visualization; Writing original draft; Writing, Review, and Editing. **Chaker Abdelaziz Kerrache:** Conceptualization; Methodology; Data Curation; Resources; Investigation; Visualization; Writing original draft; Writing, Review, and Editing. **Mohamed Chahine Ghanem:** Conceptualization; Methodology; Data Curation; Resources; Investigation; Visualization; Review, and Editing.